\documentclass[aps,prd,nofootinbib,preprint]{revtex4-1}
\usepackage{graphicx}
\usepackage{xcolor}
\usepackage{amsmath}

\def\be{\begin{equation}}
\def\ee{\end{equation}}
\def\bea{\begin{eqnarray}}
\def\eea{\end{eqnarray}}

\def\met{\not{\!\!{\rm E}}_T}
\def\ss2l{SS2$\ell$}
\def\3l{3$\ell$}

\begin{document}

\title{New LUX result compensates LHC searches for exotic quarks}
\vspace*{1cm}

\author{\vspace{1cm} Chuan-Ren Chen and  Ming-Jie Li }

\affiliation{
\vspace*{.5cm}
  \mbox {Department of Physics, National Taiwan Normal University, Taipei 116, Taiwan}\\
\vspace*{1cm}}

\begin{abstract}
The scenario of the compressed mass spectrum between heavy quark and dark matter is a challenge for LHC searches. However, the elastic scattering cross section between dark matter and nuclei in dark matter direct detection experiments can be enhanced with nearly degenerate masses between heavy quarks and dark matter. In this paper, we illustrate such scenario with a vector dark matter, using the latest result from LUX 2016. The mass constraints on heavy quarks can be more stringent than current limits from LHC, unless the coupling strength is very small. However, the compress mass spectrum with allowed tiny coupling strength makes the decay lifetime of heavy quarks longer than the time scale of QCD hadronization.  
\end{abstract}


\maketitle

\section{Introduction}
\label{sec:intro}

Even though the Standard Model (SM) has a great success in explaining all the collider experiments up to date, there are several scenarios and unanswered questions calling for new physics beyond the Standard Model (BSM), e.g. the existence of dark matter. One of the tasks of Large Hadron Collider (LHC) at CERN is searching for the signals of new particles, including dark matter, of BSM.
If new particles are charged under a $Z_2$  symmetry in BSM while the SM particles are neutral, such as $R$-parity in supersymmetry models (SUSY)~\cite{Martin:1997ns} and T-prity in Little Higgs Models~\cite{ArkaniHamed:2001nc,Cheng:2003kk},  the collider signatures of these new particles usually involve large missing transverse energy ($\met$) carried away by the undetected dark mattes. 

With data of an integrated luminosity $13~{\rm fb}^{-1}$ collected at the 13 TeV LHC, ATLAS and CMS collaborations recently updated their results. 
Using events with jets plus $\met$, the good agreement between data and SM predictions imposes the constraint of the SUSY partners of the SM  quarks  of first two generations (squark) to be heavier than $1150$ GeV with four mass-degenerate flavors, assuming that the squark decays predominately into a SM light quark and a dark matter (lightest SUSY particle, LSP)~\cite{CMS:2016mwj}. 
For the SUSY partner of top quark, called stop, the mass below $860$ GeV is excluded, assuming stop decays $100~\%$ into a top quark and a LSP~\cite{CMS:2016hxa}. 
However, these results rely on the mass of dark matter, and the limits mentioned above are based on the massless LSP hypothesis. When the mass of dark matter is close to the mass of squark (or stop), the LHC loses the detection power quickly, see Fig. 9 of Ref.~\cite{CMS:2016mwj} and Fig. 7 of Ref.~\cite{CMS:2016hxa}.
For fermonic partners of SM quarks, since the production cross sections are larger than that of squarks and stop, the constraints would be more stringent. Assuming similar behaviors of the final state kinematics, the mass limits found for squark and stop can be naively translated to fermionic quarks: the exotic quarks are bounded from below by $1485$ GeV and  $1135$ GeV for the partners of SM light quarks and for the partner of top quark, respectively.\footnote{The mass of exotic quark is given such that its pair production cross section matches that of the squark (or stop). The cross sections of pair production of squarks and exotic quarks are computed by  NLL-fast~\cite{Beenakker:2015rna} with MSTW2008NLO PDF and   HATHOR~\cite{Aliev:2010zk} with MSTW2008nnlo68cl PDF, respectively.} 

Direct search for the dark matter is performed with the detection of the signal of elastic scattering between dark matter and nuclei. Currently, no signal has been found. The null experiment result imposes constraints on scattering cross section of dark matter with a single nucleon, and can further imply the limit of  couplings between dark matter and SM quarks or gluon. 
For models with $Z_2$ parity embedded, new colored particles are usually involved in the elastic scattering processes as mediators.  These colored particles are searched at the LHC with the events of  jets or top quarks plus $\met$, assuming a benchmark decay branching ratio to dark matter and SM quarks.
The coupling strength between dark matter, heavy colored particles and SM quarks is rather irrelevant, and actually LHC has difficulty examining it. However, in the calculation of elastic scattering cross section of dark matter with nucleon, the value of the coupling strength plays a sufficient role. Furthermore, as we will see later, the elastic scattering cross section can be enhanced when the mass gap between the heavy colored particle, as a mediator, and  dark matter becomes smaller, which is the region where LHC is losing its detection power.  Therefore, experiments of direct detection of dark matter explore the coupling strength and cover the compressed mass spectrum that is difficult for LHC experiment. 
In this study, we show how the direct detection of dark matter compensates the LHC searches of heavy quarks. 
We focus on the case of a spin-1 vector dark matter and the latest result released recently by LUX experiment~\cite{Akerib:2016vxi}, which is currently the most sensitive direct search experiment for dark matter. 

The rest of this letter is organized as follows. In Section~\ref{sec:model}, we describe the relative effective Lagrangian  and give a brief review of elastic scattering between dark matter and nucleon. The numerical results are shown in Section~\ref{sect:result}. Finally, we give the summary and discussions. 

\section{Simplified model and elastic scattering}
\label{sec:model}
\begin{figure}[t]
\includegraphics[scale=0.4,clip]{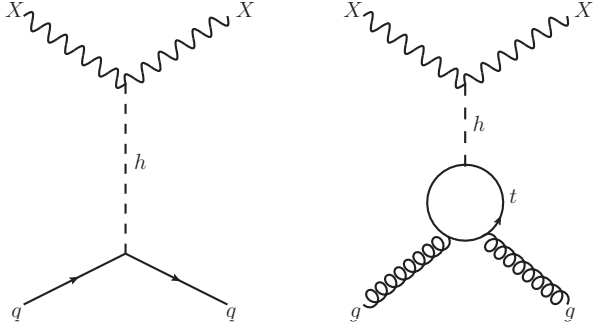}
\caption{Feynman diagram for  vector dark matter scattering from quark (left) and gluon (right) through interaction of dark matter and Higgs boson.}
\label{fig:fey1} 
\end{figure}

\begin{figure}[t]
\begin{center}
\includegraphics[scale=0.45,clip]{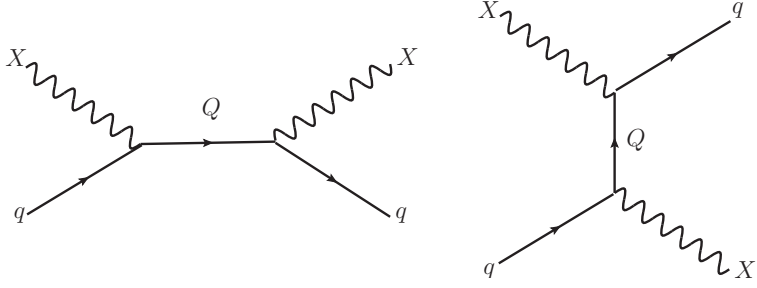}
\caption{Feynman diagrams for vector dark matter scattering with quark through new heavy quark $Q$.}
\label{fig:fey2}
\end{center}
\end{figure}

\begin{figure}[htbp]
\begin{center}
\includegraphics[scale=0.5,clip]{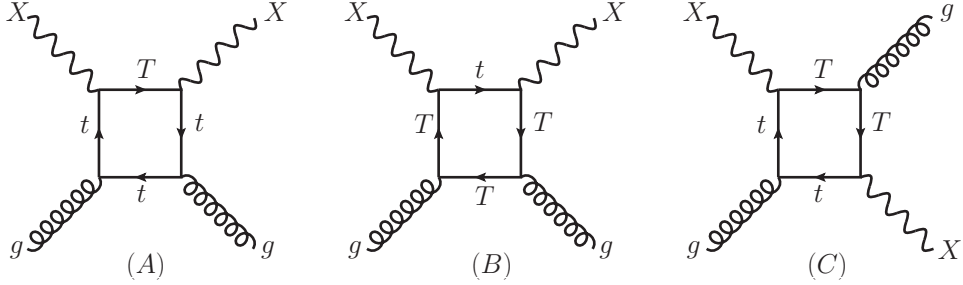}
\caption{Feynman diagrams for vector dark matter scattering with gluon through loop of new heavy quark $T$ and SM top quark $t$.}
\label{fig:fey3}
\end{center}
\end{figure}

Dark matter with spin-1 exists in  many models, such as Kaluza-Klein photon in universal extra dimension~\cite{Appelquist:2000nn} and T-odd heavy photon in Littlest Higgs model with T-parity~\cite{Cheng:2003kk}. Here we consider a simplified model where the dark matter particle is a vector boson associated with gauge symmetry $U(1)_X$~\cite{Kanemura:2010sh,Hisano:2010yh,Lebedev:2011iq,Djouadi:2011aa,Baek:2012se,Farzan:2012hh,Yu:2014pra}. The gauge invariant Lagrangian can be written as
\be
{\cal L}_{VDM} = -\frac{1}{4}X_{\mu\nu}X^{\mu\nu}+\frac{1}{2}M_{X}^2  X_\mu X^\mu +\frac{1}{4}\lambda_X(X_\mu X^\mu)^2 +\frac{1}{2}\lambda_{XH}X_\mu X^\mu H^\dagger H,
\label{eq:hvv}
\ee
where $X_\mu$ is dark matter field,  $X_{\mu\nu}=\partial_\mu X_\nu-\partial_\nu X_\mu$ is field strength tensor,  $M_X$ is mass of dark matter particle and $H$ is SM Higgs field.  The last term of the above equation induces the interaction between dark matter and Higgs boson, and  contributes to the scattering of dark matter from  gluon ($g$) through top quark loop and from light quark ($q$) at tree level, see Fig.~\ref{fig:fey1}. 
After integrating out the Higgs boson, we have the effective interactions of the vector dark matter with quarks and gluons as 
\be
{\cal L}_{eff}^{(h)} =  C_q^{(h)} m_q X^\mu X_\mu \bar{q} q + C_G^{(h)} X^\mu X_\mu G^{a\alpha\beta}G_{a\alpha\beta},
\label{eq:effh}
\ee
where
\be
 C_q^{(h)}= -\frac{\lambda_{XH}}{2m_h^2},~~C_G^{(h)}= \frac{\lambda_{XH}\alpha_s}{24\pi m_h^2}.
 \label{eq:ch}
\ee
If   the dark matter interacts with SM only through the Higgs boson, it is so-called the Higgs portal scenario (see for example~\cite{Kanemura:2010sh, Baek:2012se}). 
Here,  we consider the existence of exotic heavy quark that decays into dark matter and its SM partner quark.
The interactions between dark matter, new heavy quark ($Q$)and the SM quark ($q$) can be described in a model independent way as 
\be
{\cal L}_{DMQ} = \bar{Q} \gamma^\mu (a+b \gamma_5) q_{SM}X_\mu +h.c.
\label{eq:dmq}
\ee
where $a$ and $b$ parametrize the coupling strength. The decay width of $Q\to q X$ 
 is therefore given by
\bea
\Gamma &=& \frac{M_Q}{16\pi}\left[(a^2+b^2)\left( 1+ \frac{m_q^2}{M_Q^2}-\frac{2M_X^2}{M_Q^2}+\frac{(M_Q^2-m_q^2)^2}{M_Q^2 M_X^2}\right) -6 (a^2-b^2)\frac{m_q}{M_Q}\right] \\\nonumber
& &\lambda^{1/2}(1,m_q^2/M_Q^2,M_X^2/M_Q^2),
\label{eq:width}
\eea
where the function $\lambda(x,y,z)=x^2+y^2+z^2-2(xy+yz+xz)$. 
 Furthermore, equation~(\ref{eq:dmq}) induces the scattering of dark matter with light quark at tree level if $q=u,d,s$, see Fig.~\ref{fig:fey2}. 
 Or it initiates the scattering of dark matter with gluon through box diagrams where heavy quarks (both SM quark and its partners) flowing in the loop, as shown in Fig.~\ref{fig:fey3}, and we illustrate the case of top quark $t$ and its partner $T$ in this paper. 
%
%
%
%
 Similarly, after integrating out the mediator $Q$ in Fig.~\ref{fig:fey2} and $t$ and $T$ in the loop in Fig.~\ref{fig:fey3}, we obtain the effective Lagrangian~\cite{Hisano:2010yh}
\be
{\cal L}_{eff}^{(Q)} =  C_q^{(Q)} m_q X^\mu X_\mu \bar{q} q + C_G^{(T,t)} X^\mu X_\mu G^{a\alpha\beta}G_{a\alpha\beta} + \frac{C^{(Q)}_{AV}}{M_X}\varepsilon_{\mu\nu\rho\sigma}X^\mu i\partial^\nu B^\rho\bar q \gamma^\sigma \gamma_5 q,
\label{eq:effq}
\ee
where
\bea
\label{eq:cQ}
 C_q^{(Q)} &=& -\frac{a^2-b^2}{m_q}\frac{M_Q}{M_Q^2-M_X^2}-(a^2+b^2)\frac{M_Q^2}{2(M_Q^2-M_X^2)^2},\\\nonumber  C_G^{(T,t)} &=& \sum_{I=A,B,C}\frac{\alpha_s}{4\pi}\left[ (a^2+b^2)f^{(I)}_+(M_X, m_t, M_T) + (a^2-b^2)f_-^{(I)}(M_X, m_t, M_T)  \right],\\\nonumber
 C_{AV}^{(Q)}&=& \frac{iM_X(a^2+b^2)}{M_Q^2-M_X^2},\nonumber
 \eea
where $A,B,C$ refer to the diagrams in Fig.~\ref{fig:fey3}. For the detail of functions $f^{(I)}_{+/-}(M_X, m_t, M_T) $, we refer readers to Appendix. 

The  elastic scattering cross section of dark matter with nucleon can be expressed as the sum of "spin-independent (SI)" and "spin-dependent (SD)" cross sections  $\sigma = \sigma_{SI} + \sigma_{SD}$. In the limit of zero momentum transfer, we have the DM-nucleon scattering cross section
\bea
&&\sigma_{SI}^N=\frac{m_N^2}{\pi(M_{X}+m_N)^2}f_{N}^2\\
&&\sigma_{SD}^N=\frac{2}{\pi}\frac{m_N^2}{(M_{X}+m_N)^2}a_N^2,
 \label{eq:sisd}
\eea
where $m_N$ is the mass of nucleon (proton $p$ or neutron $n$), and $f_N$ and $a_N$ are the dark matter effective  scalar  and spin-spin interactions, respectively, to nucleon. For a spin-1 vector dark matter, we have~\cite{Hisano:2010yh}
\be
f_{p(n)} =\sum_{q=u,d,s} C_q f^{p(n)}_{Tq} + \sum_{q=u,d,s,c,b}\frac{3}{4}(q(2)+\bar q(2))C^{(2)}_q-\frac{8\pi}{9\alpha_s}C_G f_{TG},
\label{eq:fpn}
\ee 
where $C_q$ and $C_G$ are derived in equations (\ref{eq:ch}) and (\ref{eq:cQ}); $q(2)$ and $\bar q(2)$ are the second moments of parton distribution functions of quark $q$ and antiquark $\bar q$, respectively; $C^{(2)}_q$ is the coefficient of the twist-2-type coupling of dark matter and quark~\cite{Drees:1993bu}. 

Numerically, we follow the values adopted in Ref.~\cite{Hisano:2010yh}: $f^p_{T_u}=0.023$, $f^p_{T_d}=0.032$, $f^n_{T_u}=0.017$, $f^n_{T_d}=0.041$ and $f^p_{T_s}=f^n_{T_s}=0.02$. The values for second moments of parton distribution functions are given by $u(2)=0.22$, $\bar u(2)=0.034$, $d(2)=0.11$, $\bar d(2)=0.036$, $s(2)=\bar s(2)=0.026$, $c(2)=\bar c(2)=0.019$ and $b(2)=\bar b(2)=0.012$. 
For spin-spin interaction term, 
\be
a_{p(n)}=\sum_{q=u,d,s}C_{AV} \Delta^{p(n)}_q\;,
\ee
where $C_{AV}$ is shown in equation (\ref{eq:cQ}).
We take $\Delta^p_u=\Delta^n_d=0.77$, $\Delta^p_d=\Delta^n_u=-0.49$ and $\Delta^p_s=\Delta^n_s=-0.15$.

\begin{figure}[t]
\begin{center}
\includegraphics[scale=0.85,clip]{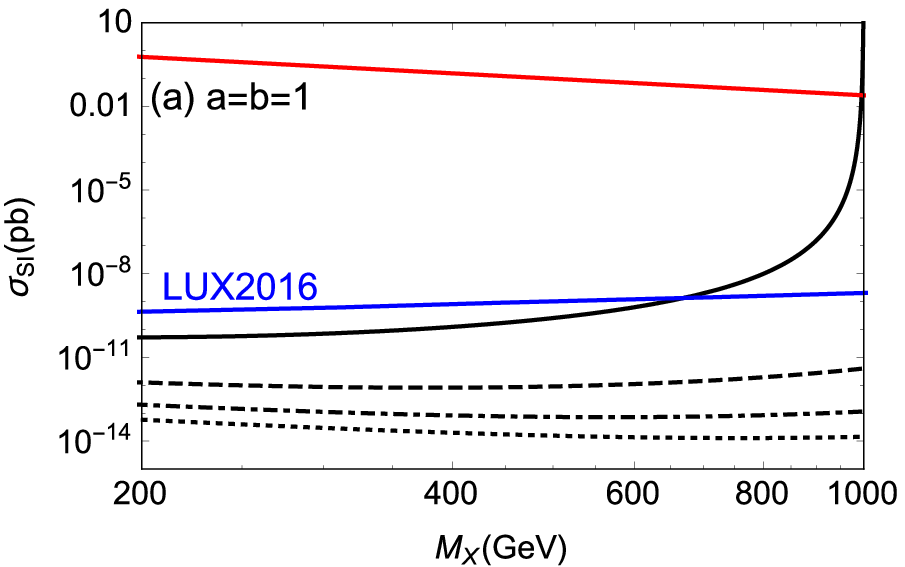}
\includegraphics[scale=0.85,clip]{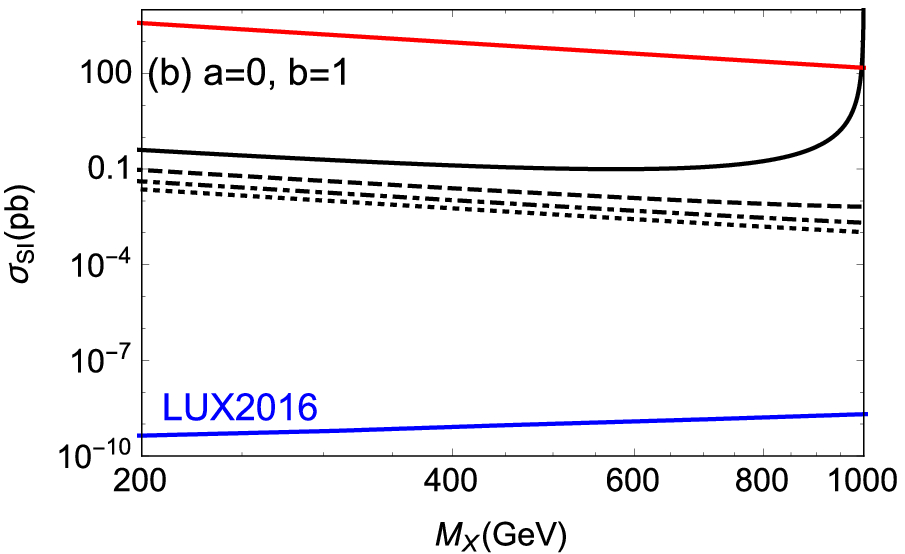}
\includegraphics[scale=0.85,clip]{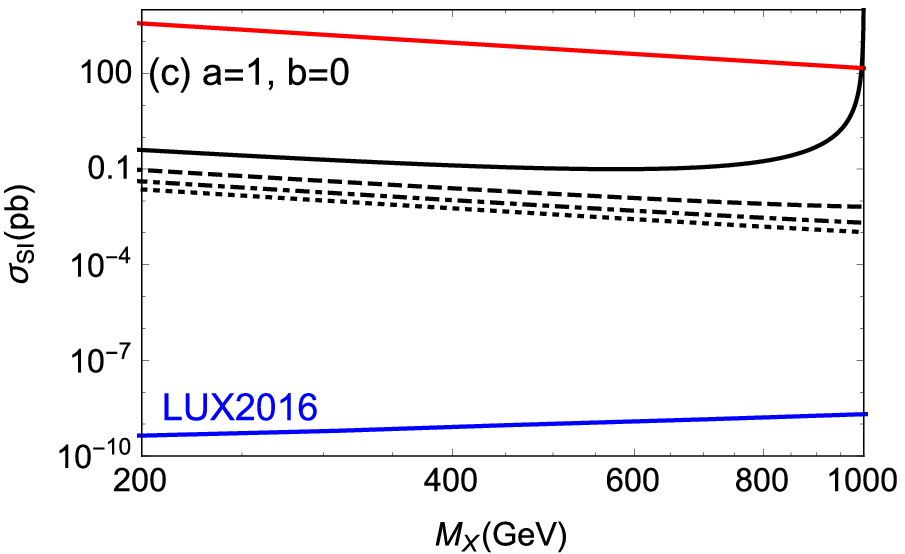}
\caption{SI elastic scattering cross section of spin-1 vector dark matter and nucleon for the interactions described in equation~(\ref{eq:dmq}). Only the heavy partners of SM up, down and strange quarks are considered. The blue curve is the current limit given by LUX~\cite{Akerib:2016vxi}. The black solid, dashed, dash-dotted and dotted curves are calculated with mass  $M_Q=1$ TeV, $2$ TeV, $3$ TeV and $4$ TeV, respectively. The case of very small mass gap between dark matter and heavy quark $M_Q-M_X=5$ GeV is shown with the red curve.}
\label{fig:sitree}
\end{center}
\end{figure}
%
\begin{figure}[htbp]
\begin{center}
\includegraphics[scale=0.29,clip]{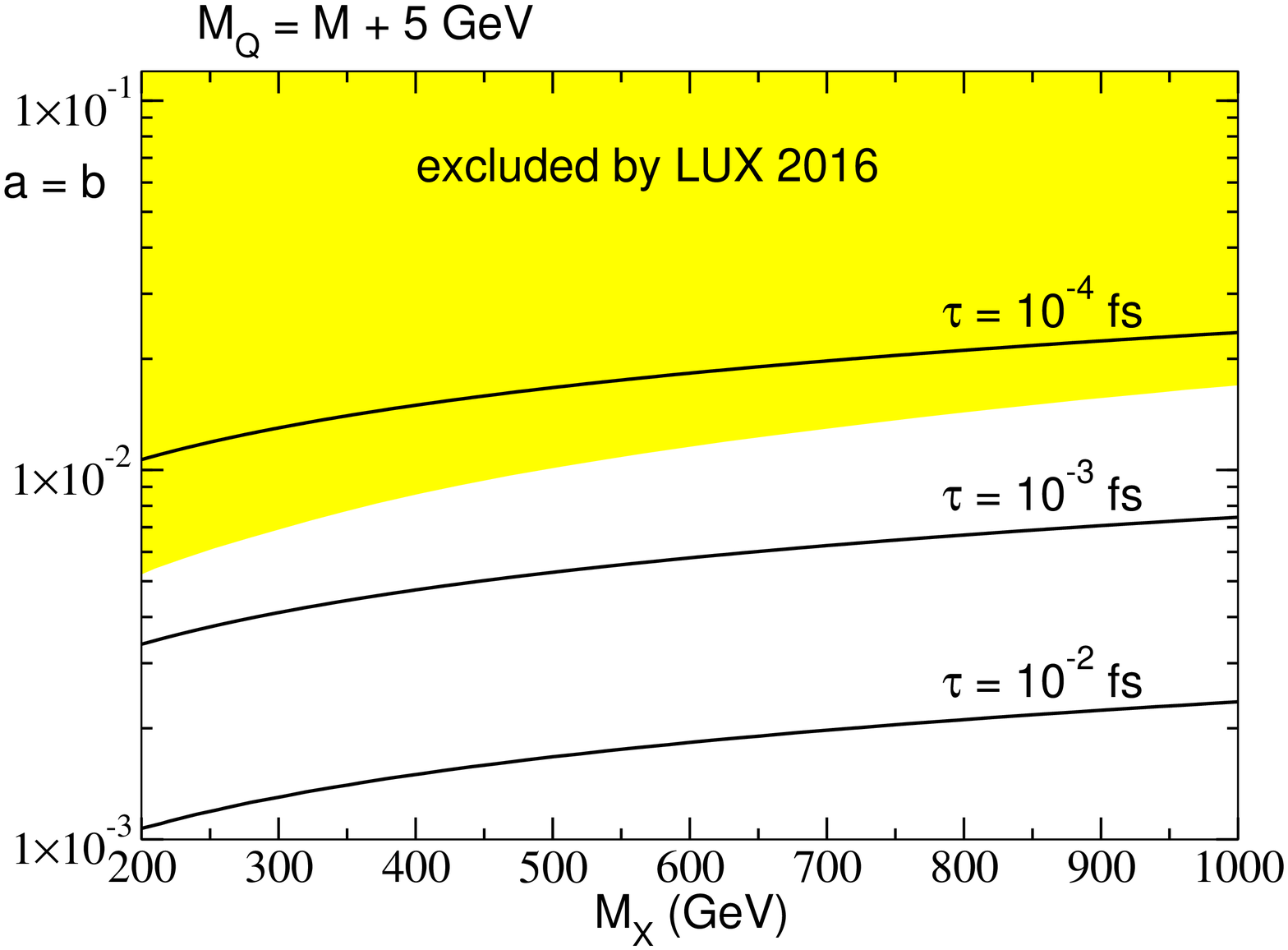}
\includegraphics[scale=0.29,clip]{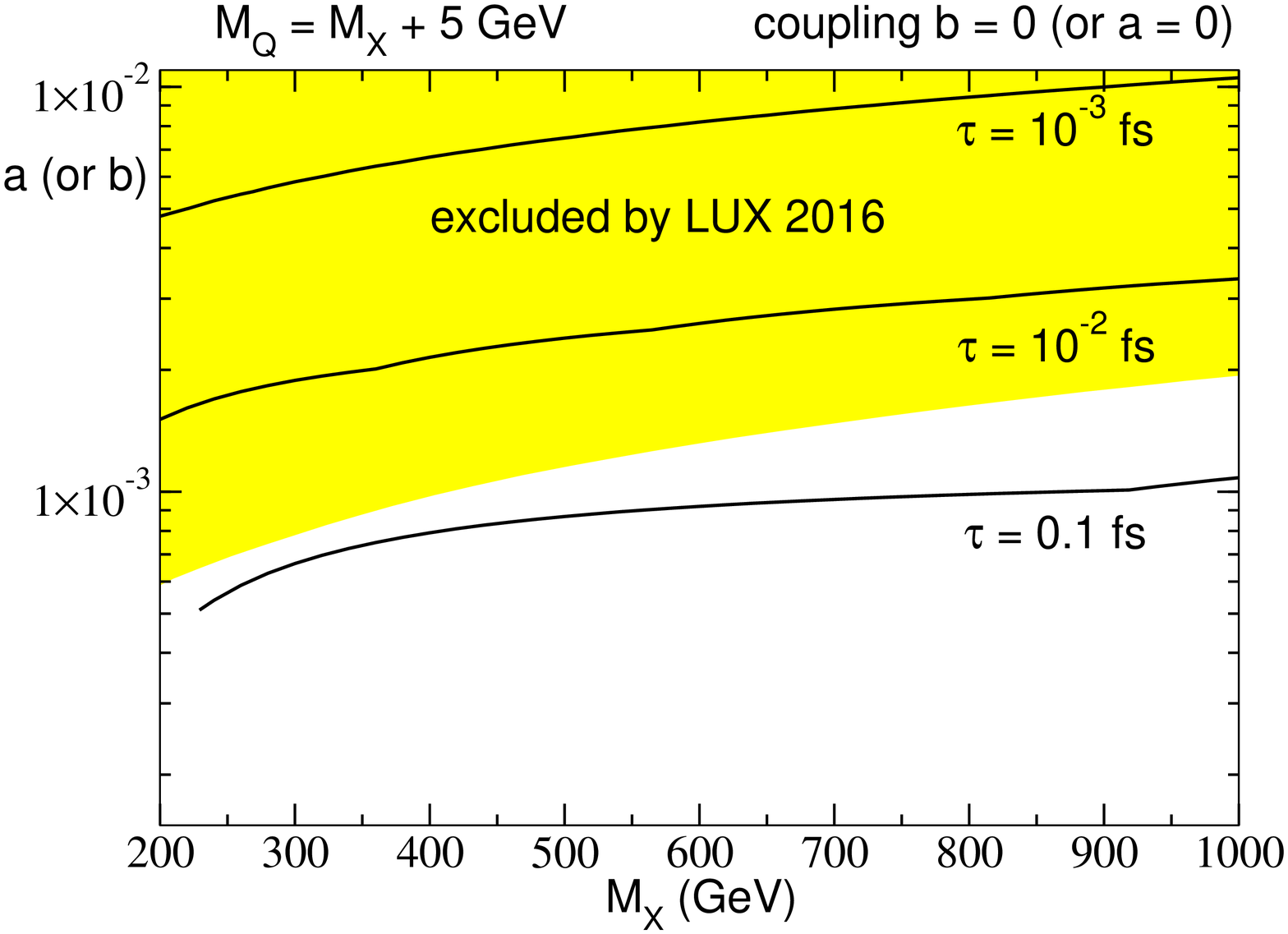}
\caption{Bound of coupling parameters $a$ and $b$ in Eq.~(\ref{eq:dmq}) from the result of LUX 2016~\cite{Akerib:2016vxi} for nearly generate spectrum $M_Q = M_X + 5$ GeV. The solid lines indicate the decay lifetime of the heavy quark $Q\to q X$. Left: $a=b$; Right: $b=0$ ($a=0$).}
\label{fig:gmtree}
\end{center}
\end{figure}
%
\section{Results}
\label{sect:result}
In this section, we begin with the numerical results of heavy quarks $Q$ that are partners of SM light quarks. We have checked that SI result is more stringent to model parameters than SD result of LUX, and we only show the comparison to $\sigma_{SI}$ of LUX through this letter. Fig.~\ref{fig:sitree} shows the SI dark matter-proton elastic scattering cross section $\sigma_{SI}$, compared with the current limit imposed by LUX 2016~\cite{Akerib:2016vxi} shown in blue curve. The cross section with $M_Q=1$ TeV,  $2$ TeV, $3$ TeV and $4$ TeV are shown in black solid, dashed, dash-dotted and dotted curves, respectively. Since $M_Q$ appears in the denominator of effective couplings in Eq.~(\ref{eq:cQ}), $\sigma_{SI}$ becomes smaller for heavier mediator $Q$. We see that all the black curves are well above the upper bound set by the LUX 2016 data, except for the case of $a=b$. 
Therefore, it is obvious that $\sigma_{SI}$ of LUX 2016 excludes heavy $Q$ with mass of several TeV, which is much more stringent than the current limit from LHC, if the coupling is purely vector ($b=0$) or axial vector ($a=0$) of ${\cal O}(1)$.
Furthermore, we also learn in the equation~(\ref{eq:effq}) that the enhancement of $\sigma_{SI}$ will occur for mass degeneracy between heavy $Q$ and dark mater, which is the case that LHC has poor sensitivity. Such a degenerate case is shown by the red curve where we take $M_Q=M_X+5$ GeV, and we can see that the LUX 2016 data excludes this compressed mass spectrum. 

\begin{figure}[htbp]
\begin{center}
\includegraphics[scale=0.85,clip]{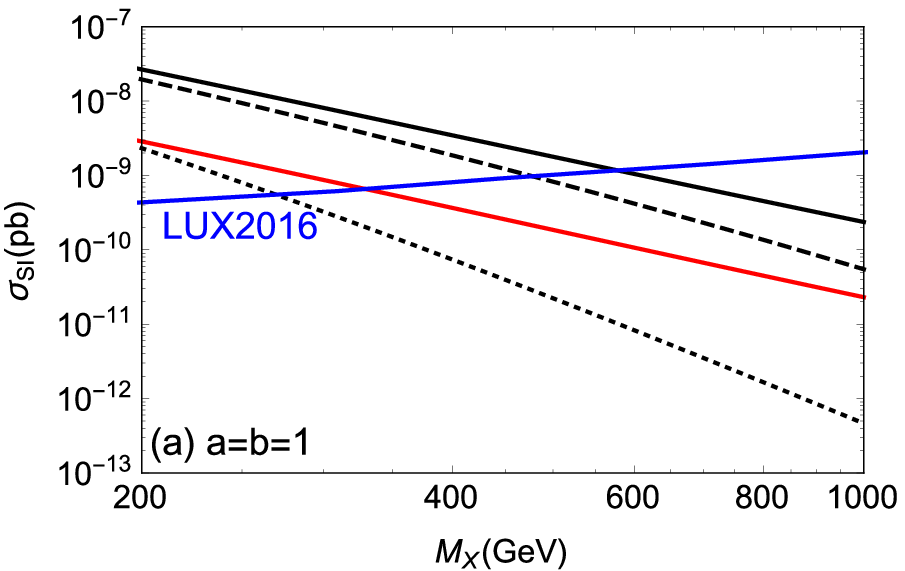}
\includegraphics[scale=0.85,clip]{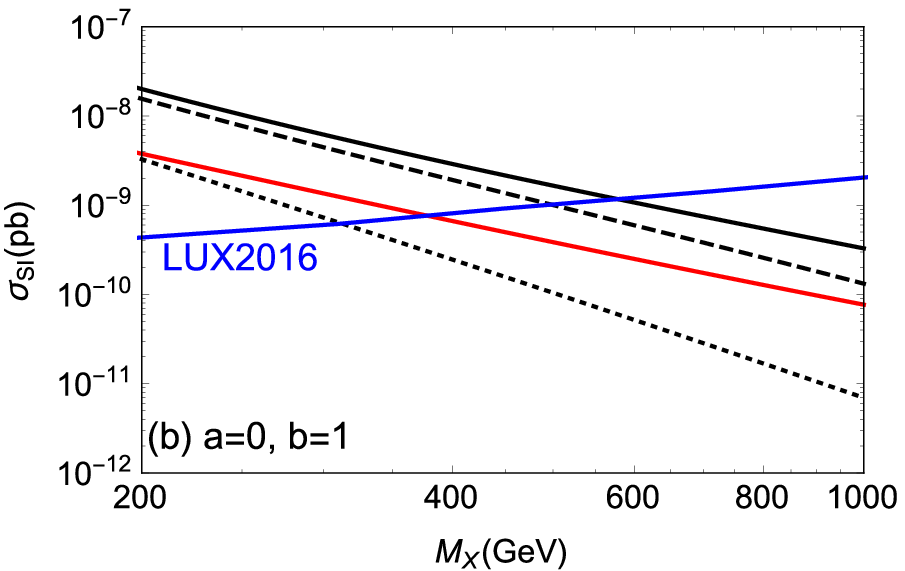}
\includegraphics[scale=0.85,clip]{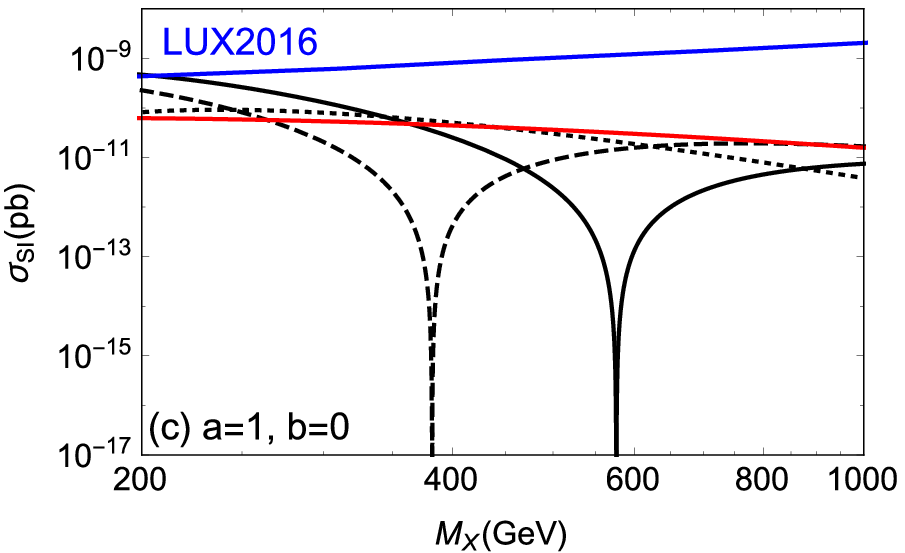}
\caption{SI elastic scattering cross section of the spin-1 vector dark matter and nucleon for the interactions described in Eq.~(\ref{eq:dmq}). Only the heavy partners of SM top quark is considered. The blue curve is the current limit given by LUX 2016~\cite{Akerib:2016vxi}, the black solid, dashed, and dotted curves are calculated with mass gap $\Delta M = M_T-M_X=0.01M_X$, $0.1M_X$ and $M_X$, respectively. The case of mass gap between heavy $T$ and dark matter $M_T-M_X = 177~{\rm GeV} \gtrsim m_t$ ($175$ GeV) is shown with the red curve.}
\label{fig:sibox}
\end{center}
\end{figure}
%
Fig.~\ref{fig:gmtree} shows the parameter space of coupling strength and the mass of dark matter that is constrained by the LUX 2016 when heavy quark $Q$ and dark matter $X$ are nearly degenerate.  Roughly speaking, the coupling $a=b \gtrsim 10^{-2}$ is excluded, see the yellow shaded region. In the case of vector coupling $b=0$ (or axial vector coupling, $a=0$), the lower bound is one order of magnitude stronger, LUX excludes the region of $a \gtrsim 10^{-3}$ ($b\gtrsim 10^{-3}$). Also shown is the decay lifetime of the heavy quark $Q\to q X$. Note that the time scale of hadronization can be estimated as $\tau_{had}\simeq 1/\Lambda_{QCD} \simeq 3\times10^{-9}$ fs with $\Lambda_{QCD}\simeq 200$ MeV~\cite{Grossman:2008qh}. Since the lifetime of heavy quark is longer than $\tau_{had}$, it is likely that the heavy quark will form a bound state before decaying into a SM light quark and dark matter. The search for the heavy quark in this parameter is challenging and interesting~\cite{Grossman:2008qh}, and is beyond the scope of our study in this letter. The degenerate case of dark matter and heavy quark particles has also been studied in literature~\cite{Hisano:2011um,Asano:2011ik,Chen:2014wua}

Now we turn to the case of heavy $T$ that interacts with SM top quark $t$ and dark matter. 
The  SI elastic dark matter-proton scattering cross sections are shown in Fig.~\ref{fig:sibox} with ${\cal O}(1)$ coupling strength.  
The black solid, dashed, dotted and red curves are the $\sigma_{SI}$ with $M_T$ being $1~\%$, $10~\%$, $100~\%$ and $177$ GeV, respectively,  heavier than mass of dark matter $M_X$. 
In the limit of $M_T \gg m_t$, the mass difference $M_T^2-M_X^2$ appears in the denominator of functions $f^{(I)}_{\pm}(M_X,m_t,M_T)$~\cite{Hisano:2010yh}, therefore, it can be understood that a small mass gap between heavy $T$ and dark matter will enhance  $\sigma_{SI}$. 
We see that, for the case of  $a=0,~b=1$ in Fig.~\ref{fig:sibox}(b), the mass of heavy $T$ quark  is constrained to be heavier than about $600$ GeV. For $a=b=1$ in Fig.~\ref{fig:sibox}(a), the bound is slightly lower. If we translate this limit to a scalar colored top partner, the lower mass bound is about $450$ GeV.
In the case of $a=1,~b=0$ (purely vector coupling) shown in Fig.~\ref{fig:sibox}(c), no constrain can be established. The deep structure of $\sigma_{SI}$ for very small mass gap $\Delta M/M_X = 1\%$ and $\Delta M/M_X = 10\%$, where $\Delta M = M_T-M_X$, is because of a cancellation happening between  $f_+^{(I)}$ and $f_-^{(I)}$ in $C_G^{(T,t)}$  in the equation (\ref{eq:cQ}). 

\begin{figure}[htbp]
\begin{center}
\includegraphics[scale=0.29,clip]{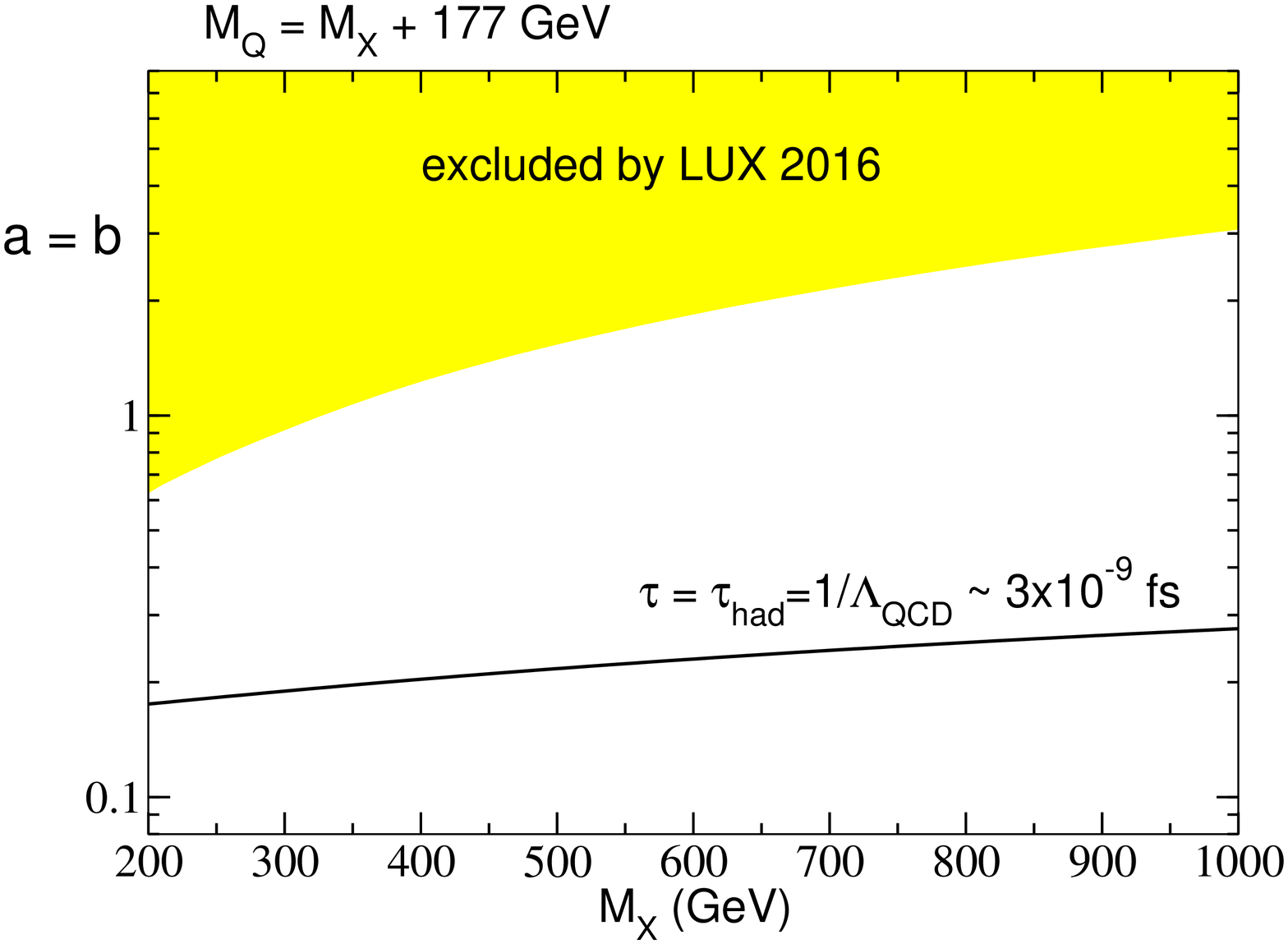}
\includegraphics[scale=0.29,clip]{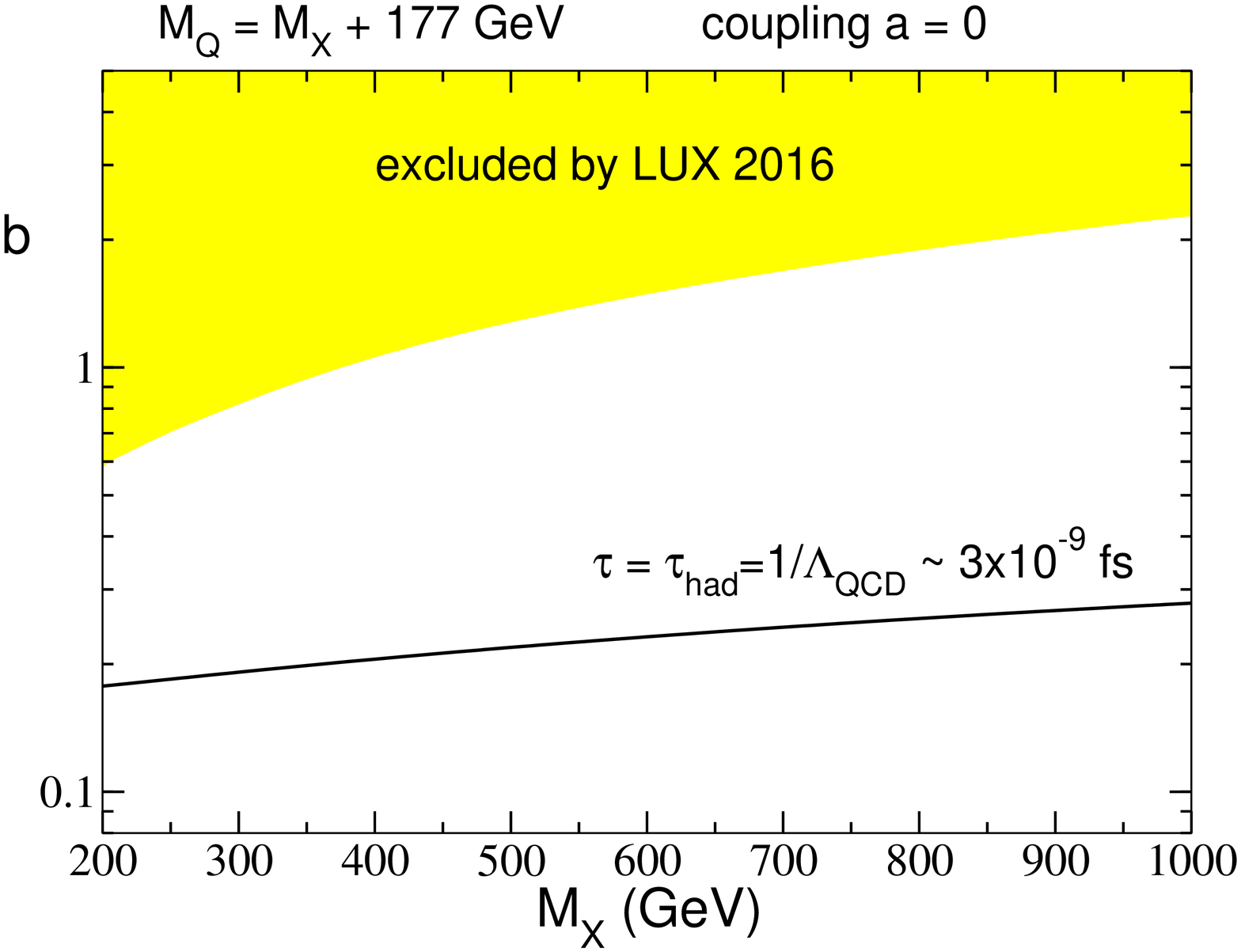}
\includegraphics[scale=0.29,clip]{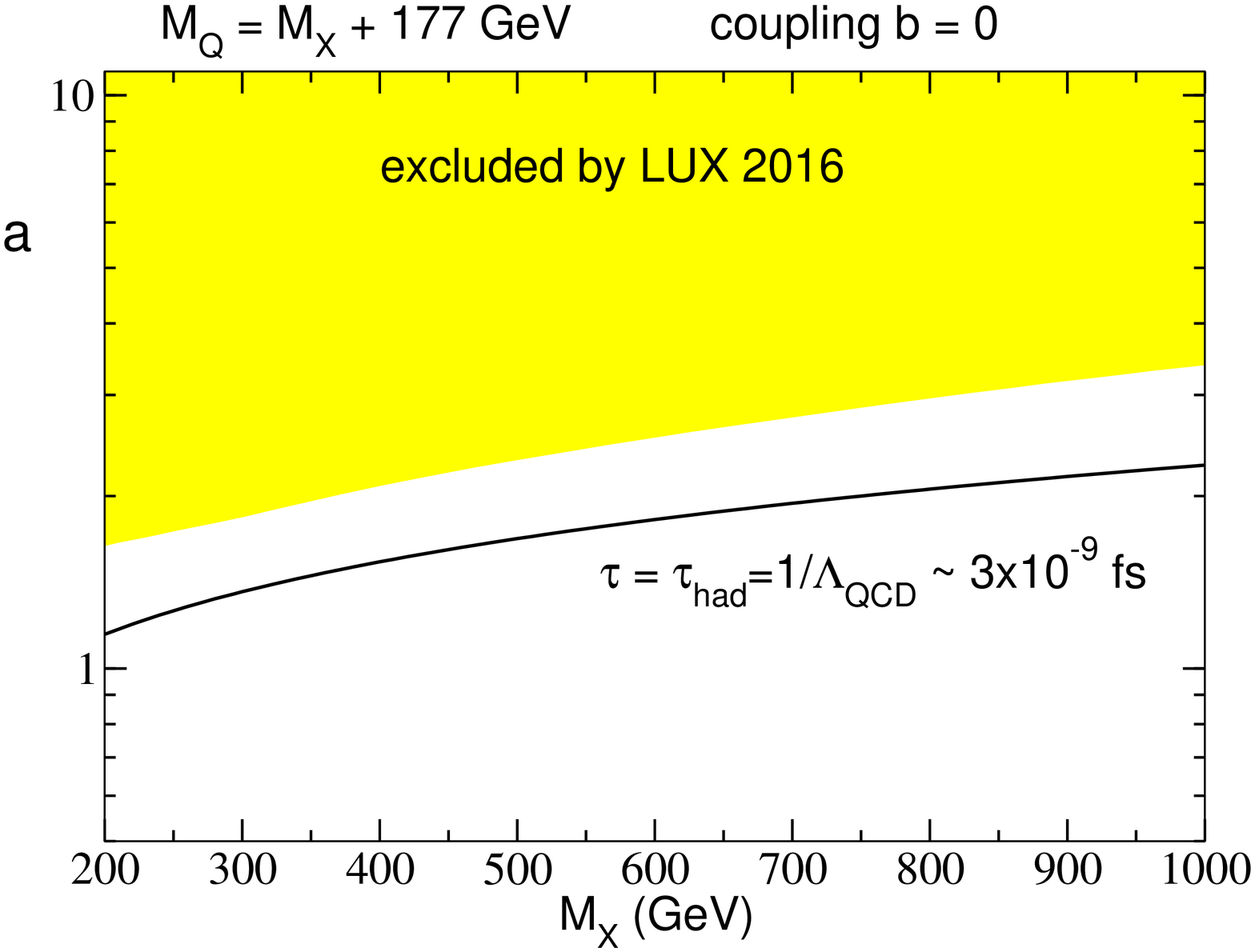}
\caption{Bound of coupling parameters $a$ and $b$ in equation~(\ref{eq:dmq}) from the result of LUX 2016~\cite{Akerib:2016vxi} for mass  spectrum $M_T \gtrsim M_X + m_t$ GeV. The solid line indicates decay lifetime of $T\to t X$ that maches the time scale for the hadronization to take place.  Top-Left: $a=b$; Top-Right: $a=0$; Bottom: $b=0$.}
\label{fig:gmbox}
\end{center}
\end{figure}

When the heavy $T$ decays into a top quark and a dark matter with $M_T \gtrsim m_t+M_X$, it is a challenge for the LHC to search for, due to the soft top quark in the decay product. 
In Fig.~\ref{fig:gmbox} we also show  the constrains of the coupling between heavy $T$ to top quark and dark matter, taking $M_T=M_X + 177$ GeV for illustration. The yellow region is excluded by the LUX 2016 data. We also show the decay lifetime of heavy $T$ that matches the time scale of QCD hadronization with black solid line, below which the decay of $T\to t X$ takes place longer than the time scale of  hadronization.  At this parameter space, heavy $T$ can form a bound state after being  produced in pair at the LHC. The detailed phenomenology study is beyond the scope of this paper and is left for a future study. 

%

\section{Summary}
\label{sect:outlook}

LHC searches for heavy quark that decays predominately into a SM quark and a dark matter face two difficulties. One is the parameter where heavy quark and dark matter are nearly degenerate. The other is the  poor capability to explore the coupling strength of heavy quark to SM quark and dark matter. 
In this paper, we point out that the dark matter direct search experiments compensate the insensitive region for LHC collider experiment, 
since the elastic scattering cross section between dark matter and nucleon is enhanced with the degenerate mass spectrum and depends significantly on coupling strength. 
We study two cases: the heavy quark is tha partner of SM (1) light quarks (u, d, s) (2) top quark. 

For the heavy quarks $Q$ that couple to SM light quarks and dark matter, with ${\cal O}(1)$ coupling strength, the lower mass bound given by the latest data of LUX 2016 can be much more stringent  than the LHC direct searches. 
Furthermore, the degenerate region is excluded, unless the coupling strength is smaller than ${\cal O}(10^{-2})$. 
However, such a tiny coupling strength brings an interesting possibility that heavy quark $Q$ will form  a bound state since its decay lifetime of $Q\to q X$ is much longer than the  time scale of QCD  hadroniztion. 
For the heavy top quark partner $T$, LUX 2016 requires $M_T > 600$ GeV for the degenerate mass spectrum with ${\cal O}(1)$ coupling strength, unless the coupling is purely vector coupling. When $M_T\simeq M_X+m_t$, in the case of purely vector coupling, LUX 2016 excludes the coupling strength larger than about 1.6 for $m_X=200$ GeV and about $3.4$ for $m_X=1000$ GeV. It is interesting to note that, as coupling strength is smaller than about $1.1$ for $m_X=200$ GeV (about $2.3$ for $m_X=1000$ GeV), the decay lifetime of $T\to t X$ is larger than the time scale of hadronization.

In summary, we see that the compressed mass spectrum of dark matter and heavy quark, which is a challenge for LHC searches for heavy quarks, can be explored with direct dark matter search experiments. The allowed region from LUX 2016 result brings an interesting issue of bound states of heavy quarks. 
 
\begin{acknowledgments}
C.-R.~C. would like to acknowledge the support of National Center for Theoretical Sciences (NCTS). This work  is supported in part by the Mistry of Science and Technology (MOST), R.O.C. under Grants No.~MOST102-2112-M-003-001-MY3 and MOST105-2112-M-003-010-MY3.%
\end{acknowledgments}

\appendix
\section{$f^{(I)}_\pm(M_X, m_t, M_T) $}
For completeness, we list the formula of functions $f^{(I)}_\pm(M_X, m_t, M_T) $~\cite{Hisano:2010yh}:
\bea
&&f^{(A)}_+ (M_X, m_t, M_T)  =\\\nonumber
&&  \frac{1}{6\Delta^2 M_X^2}\left[ \Delta \left( M_X^2 (M_T^2-m_t^2) + m_t^2 (m_t^2 + 5M_T^2)\right) - 6 m_t^2 M_T^2 \left( (M_T^2 - m_t^2)^2 - M_X^2 (m_t^2 +3 M_T^2) \right) \right] \\\nonumber
&&- \frac{m_t^2}{12 M_X^2}\ln\left( \frac{m_t^2}{M_T^2} \right) \\\nonumber
&&+ \frac{m_t^2 L}{12 \Delta^2 M_X^4} [ (M_T^2+m_t^2-M_X^2)\Delta^2 + 2 M_T^2\Delta\left(5M_T^4 + 20m_t^2M_T^2-m_t^4 + M_X^2(9M_T^2+m_t^2)\right)\\\nonumber
&&~~~~~~~~~~~~~+12M_T^4\left(M_X^2(M_T^4+10m_t^2M_T^2+5M_t^4)-(M_T^2-m_t^2)^2(m_T^2+2m_t^2) \right)].\\\nonumber
 \label{eq:fap}
 \eea
\bea
f^{(A)}_- (M_X, m_t, M_T)  &=& -\frac{M_T}{6m_t \Delta^2}\left[ \Delta(2M_T^2 + m_t^2 -2M_X^2)+6m_t^2M_T^2(M_T^2-m_t^2-M_X^2) \right]\\\nonumber
&& +\frac{L}{\Delta^2}m_t M_T^3\left[  \Delta+m_t^2(M_T^2 - m_t^2 + M_X^2) \right]
 \label{eq:fam}
 \eea
\bea
f^{(B)}_+ (M_X, m_t, M_T)  &=& f^{(A)}_+ (M_X, m_t \to M_T, M_T\to m_t) \\\nonumber
f^{(B)}_- (M_X, m_t, M_T)  &=& f^{(A)}_+ (M_X, m_t \to M_T, M_T\to m_t) 
 \label{eq:fb}
 \eea
\bea
&&f^{(C)}_+ (M_X, m_t, M_T)  = -\frac{1}{6\Delta M_X^2}\left[ 2M_X^4 - 3M_X^2(m_t^2+M_T^2) + (m_t^2 - M_T^2)^2\right] \\\nonumber
&& + \frac{m_t^2}{12 M_X^4}\ln\left(\frac{m_t^2}{M_T^2}\right)+  \frac{M_T^2}{12 M_X^4}\ln\left(\frac{M_T^2}{m_t^2}\right)\\\nonumber
&& + \frac{L}{12\Delta M_X^4} \left[ \Delta(M_X^2 -m_t^2-M_T^2)(m_t^2+M_T^2)+4m_t^2M_T^2[(m_t^2-m_T^2)^2-2M_X^2(m_t^2+M_T^2)] \right] \\\nonumber
&&f^{(C)}_- (M_X, m_t, M_T)  = 0
 \label{eq:fc}
 \eea
In the above equations, the definitions of $\Delta$ and $L$ are:
\be
\Delta = M_X^4 - 2M_X^2(m_t^2+M_T^2) + (M_T^2-m_t^2)^2,\\
 \label{eq:delta}
 \ee
\bea
L &&=\frac{1}{\sqrt{|\Delta|}}\ln\left( \frac{M_T^2+m_t^2-M_X^2+\sqrt{|\Delta|}}{M_T^2+m_t^2-M_X^2-\sqrt{|\Delta|}} \right)~{\rm for}~\Delta>0;\\
L &&=\frac{2}{\sqrt{|\Delta|}}\tan^{-1}\left( \frac{\sqrt{|\Delta|}}{M_T^2+m_t^2-M_X^2} \right)~{\rm for}~\Delta<0.
 \label{eq:LL}
 \eea

%

\end{document}